\documentclass[utf8]{frontiersFPHY} % for Physics and Applied Mathematics and Statistics articles

\usepackage{url,hyperref,lineno,microtype,subcaption}
\usepackage[onehalfspacing]{setspace}
\usepackage[T1]{fontenc} 
\usepackage{color}
\def\keyFont{\fontsize{8}{11}\helveticabold }

\def\firstAuthorLast{Andersen {et~al.}} %use et al only if is more than 1 author
\def\Authors{Brian M. Andersen$^{1}$, Andreas Kreisel$^{1}$, and P. J. Hirschfeld$^{2}$}
\def\Address{$^{1}$Niels Bohr Institute, University of Copenhagen, DK-2200 Copenhagen, Denmark\\
$^{2}$Department of Physics, University of Florida, Gainesville, Florida 32611, USA}
\def\corrAuthor{Brian M. Andersen}

\def\corrEmail{bma@nbi.ku.dk}

\definecolor{blue}{rgb}{  0,  0,    1}
\definecolor{d4blue}{rgb}{  0,  0.4470,    0.7410}
\definecolor{d12orange}{rgb}{0.8500    0.3250    0.0980}
\definecolor{g8yellow}{rgb}{0.    0.6    0.298}
\definecolor{ppurple}{rgb}{0.4940    0.1840    0.5560}

\begin{document}
\onecolumn
\firstpage{1}

\title[]{Spontaneous time-reversal symmetry breaking by disorder in superconductors} 

\author[\firstAuthorLast ]{\Authors} %This field will be automatically populated
\address{} %This field will be automatically populated
\correspondance{} %This field will be automatically populated

\extraAuth{}% If there are more than 1 corresponding author, comment this line and uncomment the next one.
%\extraAuth{corresponding Author2 \\ Laboratory X2, Institute X2, Department X2, Organization X2, Street X2, City X2 , State XX2 (only USA, Canada and Australia), Zip Code2, X2 Country X2, email2@uni2.edu}

\maketitle
\begin{abstract}
 A growing number of superconducting materials display evidence for spontaneous time-reversal symmetry breaking (TRSB) below their critical transition temperatures. Precisely what this implies for the nature of the superconducting ground state of such materials, however, is often not straightforward to infer. We review the experimental status and survey different theoretical mechanisms for the generation of TRSB in superconductors. In cases where a TRSB complex combination of two superconducting order parameter components is realized, defects, dislocations and sample edges may generate superflow patterns that can be picked up by magnetic probes. However, even single-component condensates that do not break time-reversal symmetry in their pure bulk phases can also support signatures of magnetism inside the superconducting state. This includes, for example, the generation of localized orbital current patterns or spin-polarization near atomic-scale impurities, twin boundaries and other defects. Signals of TRSB may also arise from a superconductivity-enhanced Ruderman-Kittel-Kasuya-Yosida exchange coupling  between magnetic impurity moments present in the normal state. We discuss the relevance of these different mechanisms for TRSB in light of recent experiments on superconducting materials of current interest.
\tiny
 \keyFont{ \section{Keywords:} Time-reversal symmetry breaking, superconductivity, disorder, condensed matter theory, quantum materials.} %All article types: you may provide up to 8 keywords; at least 5 are mandatory.
\end{abstract}

\section{Introduction}

Several superconducting materials have been discovered to manifest evidence of spontaneous time-reversal symmetry breaking (TRSB) that appears only below their critical superconducting transition temperatures $T_{\mathrm{c}}$~\cite{Kallin16,Wysokinski2019,Ghosh2EA20,Henrik2022}. Evidence for such TRSB originates mainly from an enhanced muon-spin relaxation ($\mu$SR) and/or polar Kerr effect measurements finding a change in the optical polar Kerr angle below $T_{\mathrm{c}}$. Additional probes able to detect breaking of time-reversal symmetry in the superconducting phase include Josephson interferometry~\cite{Strand2009}, superconducting quantum interference device (SQUID) magnetometry~\cite{HicksEA10}, and polarized and small-angle neutron scattering~\cite{Avers2020}. These experimental methods detect the internal magnetic fields spontaneously generated
in the superconducting state.
Explanations for the occurrence of such fields from the perspective of clean (homogeneous) physical effects  are either in terms of a TRSB  non-unitary spin-triplet pairing state or in terms of multi-component superconducting condensates entering a complex TRSB superposition of two superconducting order parameter components. The latter option is particularly natural for superconducting instabilities condensing in two-dimensional (2D) irreducible representations of the associated crystal point group. Complex superpositions of two symmetry-distinct order parameters break TRSB, and generate persistent supercurrent patterns at material edges, dislocations, or around various defect sites, which produce magnetic fields that may not average to zero locally. Such a multi-component scenario for the origin of TRSB below $T_{\mathrm{c}}$ has been extensively discussed for e.g. Sr$_2$RuO$_4$~\cite{Henrik2022,MaenoEA94, MackenzieEA17,RomerEA19,RomerEA21,KivelsonEA20}, UPt$_3$~\cite{StewartEA84,JoyntEA02}, URu$_2$Si$_2$~\cite{Kawasaki2014}, UTe$_2$~\cite{RanEA19,AokiEA22}, SrPtAs~\cite{Biswas2013}, Re~\cite{Shang2018}, Zr$_3$Ir~\cite{Shang2020}, LaNiC$_2$~\cite{HillierEA09}, LaNiGa$_2$~\cite{Hillier2012},  PrOs$_4$Sb$_{12}$~\cite{BauerEA02,AokiEA03}, Ba$_{1-x}$K$_x$Fe$_2$As$_2$~\cite{RotterEA08,BokerEA17}, FeSe$_{1-x}$S$_x$\cite{Matsuura2023}, and more recently also for the kagome superconductors $A$V$_3$Sb$_5$ ($A$: K, Rb, Cs)~\cite{Guguchia2022Tunable,Mielke2022Time-reversal,Romer2022}, 4Hb-TaS$_2$~\cite{Ribak2020}, {CaPd}$_{2}${Ge}$_{2}$~\cite{Anand2023}, skutterudites~\cite{Kataria2023,Kataria2023a} and {Na}$_2${Cr}$_3${As}$_3$ \cite{Bhattacharyya2023}. For most of the compounds listed here, the evidence for TRSB originates from $\mu$SR experiments. For further materials displaying TRSB in their superconducting phases, and other types of evidence, we refer to Refs.~\cite{Kallin16,Wysokinski2019,Ghosh2EA20,Henrik2022}.

At present, the status of the precise superconducting ground state remains controversial for many of the materials listed above. This is the case, for example, for the two materials Sr$_2$RuO$_4$ and UTe$_2$, where specific heat data features only a single thermodynamic transition that does not split under uniaxial strain, casting considerable doubt on the multi-component nature of their superconducting condensates~\cite{SteppkeEA17,LiEA19,RosaEA22,GirodEA22}. Therefore, it appears timely and important to pursue other possibilities for the origin of TRSB detected solely inside the superconducting phase. This motivates the question of what mechanisms there are for TRSB of superconducting condensates composed of a single component order parameter? Barring non-unitary TRSB spin-triplet order, the answer to this question naturally leads to a study of different forms of spatial inhomogeneities and their influence on the local properties of superconductivity. For example, it is known that some superconductors generate magnetic moments below $T_{\mathrm{c}}$ due to an interplay between the gap structure and electronic correlations~\cite{Tsuchiura2001,ZWang2002,Zhu2002,Chen2004,Andersen2007,Harter2007,Andersen2007,Andersen2010}. As mentioned above, another possibility for TRSB appearing only below $T_{\mathrm{c}}$ includes the generation of localized orbital loops of supercurrents near impurities. The latter is well-known to arise near nonmagnetic disorder sites in complex TRSB multi-component condensates~\cite{Lee2009,Garaud2014,Maiti2015,Lin2016,GaraudPRL,Benfenati,Merce2022,Yerin2022}, but was recently shown to be also present in the strongly disordered regime of single-component superconductors~\cite{Li2021,Clara2022}. Likewise, one might expect dislocations and grain boundaries to similarly operate as seeds of localized supercurrents producing sizable internal magnetic fields. Recently, the latter scenario was studied in Refs.~\cite{WillaEA21} and \cite{Clara2023} as a possible explanation for reconciling the existence of TRSB and a single specific heat transition in Sr$_2$RuO$_4$. Indeed for this material, edge dislocations are known to be prevalent in many samples~\cite{Ying2013}. Finally, one might also envision that time-reversal symmetry is already broken above $T_{\mathrm{c}}$ through the existence of small magnetic regions that are either too dilute or rapidly fluctuating to be picked up by most probes. Only through a superconductivity-enhanced coupling, or a sufficient slowing-down of such preexisting magnetism below $T_{\mathrm{c}}$, can it be detected, e.g. by entering the muon relaxation time window.

Here, motivated by the above-mentioned developments we survey a number of different mechanisms for {\it inhomogeneity-induced}  TRSB in superconductors. Our discussion of disorder-induced TRSB refers both to cases where TRSB exists only locally near inhomogeneity sites and where TRS is broken globally, but manifested via induced currents near disorder sites. We discuss these both in view of recent theoretical and experimental developments of unconventional superconductivity. 

\section{Mechanisms of TRSB from disorder in superconductors }

In this section we discuss three different mechanisms for disorder-generated TRSB in spin-singlet superconductors. Section~\ref{complexmechanism} reviews the generation of magnetic fields from disorder in superconducting phases with TRSB in the bulk or generated near spatial inhomogeneities. Section~\ref{momentmechanism} discusses another case where disorder in conjunction with electronic interactions and appropriate superconducting gap structures lead to the generation of static magnetic moments. Due to the required gapping of states for these moments to form, this mechanism is tied to the superconducting phase. Finally, Sec.~\ref{couplingmechanism} surveys a third case where preexisting magnetic local order (fluctuation moments) in the normal state get enhanced (slowed-down) by superconductivity. Clearly these mechanisms do not exhaust the options for disorder-induced TRSB in superconductors where e.g. cases with substantial spin-orbit coupling may also be of interest~\cite{GrafBalatsky00,Hano23}. We focus on these three options given recent research along similar lines and for their relevance for some of the superconducting materials mentioned above. For a discussion of the possibility of TRSB in homogeneous superconductors originating from conventional electron-phonon coupling, we refer to the review~\cite{Ghosh2EA20} and the references therein. A theoretical proposal for a mechanism has been proposed in Ref.~\cite{Agterberg1999} and recent experimental evidence for TRSB in materials where superconductivity is believed to originate from electron-phonon coupling is reported in Refs.~\cite{Bhattacharyya2022,Anand2023,Kataria2023,Kataria2023a}.

\subsection{Complex combinations of two superconducting condensates}
\label{complexmechanism}

\begin{figure*}[tb]
    \centering
    \includegraphics[width=1.0\linewidth]{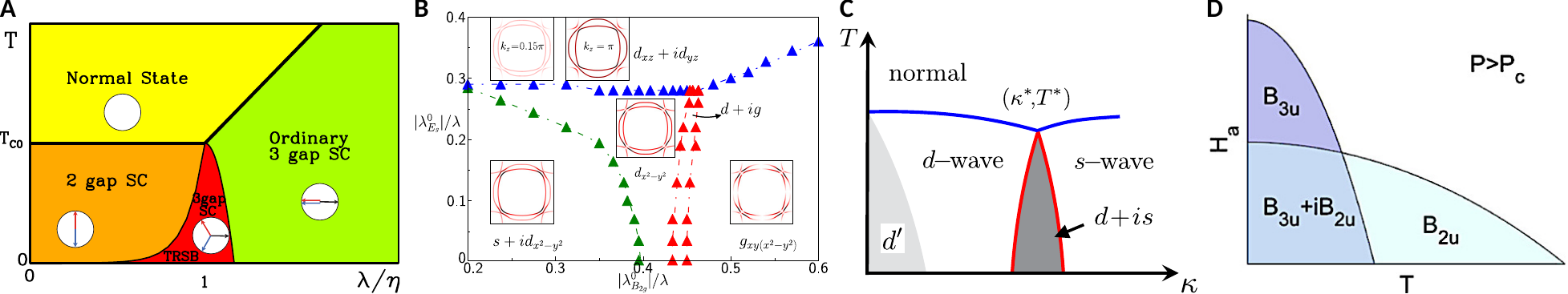}
    \caption{Selection of theoretical proposals for phase diagrams exhibiting TRSB in homogeneous superconducting phases. {\bf A} Frustrated multiband superconductor which exhibits, as a function of frustration parameter, a state with complex order parameters of $s+is$ type. The three order parameters are symbolized as arrows in the white circles where the magnitude is the arrow length while the direction indicates phase in the complex plane~\cite{Stanev2010,Carlstroem2011,Maiti2013}.
     {\bf B} Phase diagram relevant for Sr$_2$RuO$_4$ exhibiting a TRSB phase of degenerate solutions       $d_{xz}+id_{yz}$~\cite{Clepkens2021}.
     {\bf C} Phase diagram calculated for Fe-based systems with only electron pockets which support $d$-wave instabilities and $s$-wave states; if these become degenerate (almost degenerate) as a function of a tuning parameter $\kappa$, a TRSB superconducting state is stabilized at and below $T_{\mathrm{c}}$~\cite{Khodas2012}.
     {\bf  D} Qualitative magnetic field-temperature phase diagram proposed for UTe$_2$ where accidentally degenerate (nearly degenerate) triplet states produce a TRSB state at $T_{\mathrm{c}}$ (below $T_{\mathrm{c}}$)~\cite{Shishidou2021}.
     (A) Reprinted figure with permission from \cite{Stanev2010}, Copyright 2010 by the American
Physical Society.
(B) Reproduced from \cite{Clepkens2021}. CC BY 4.0.
(C)  Reprinted figure with permission from \cite{Khodas2012}, Copyright 2012 by the American Physical Society.
(D) Reprinted figure with permission from \cite{Shishidou2021}, Copyright 2021 by the American
Physical Society.}
    \label{fig:fig0}
\end{figure*}

Entering the superconducting state is associated with a breaking of the symmetry of the normal state symmetry group composed of a direct product of crystalline symmetries, gauge symmetry and the time reversal operation~\cite{Volovik1985,Annett90,SigristUeda91}.
For non-magnetic normal states, the allowed solutions of the gap equation can be characterized by the irreducible representations (irreps) of the associated point group. The leading instability is  found from the linearized BCS gap equation as the solution with the largest eigenvalue and thereby also the highest $T_{\mathrm{c}}$~\cite{Graser2009,Romer2015,Roemerweaktostrong,Kreisel_review}.
Broken time-reversal symmetry can arise from a complex combination of superconducting condensates or the breaking of the spin-rotational invariance in non-unitary triplet states.
Whenever the preferred superconducting instability corresponds to a two-dimensional (2D) irrep, the system may take advantage of the degeneracy between the representations. From a perspective of the minimization of the Ginzburg-Landau free energy, there is the possibility to either form a real combination thereby breaking a crystal symmetry or to form complex superpositions~\cite{Ghosh2EA20,KivelsonEA20,Annett90}. The latter often maximizes the condensation energy by forming a fully-gapped state and can be shown to be the ground state for a single band setting within a loop expansion~\cite{Roising2023}. We will not elaborate further on the possibility of real superpositions, but  focus instead on complex linear combinations with standard examples as $p_x+ip_y$ ($p+ip$) chiral order ($E_u$) in tetragonal systems and $d_{xy}+id_{x^2-y^2}$ ($d+id$) ordered superconductivity ($E_2$) in hexagonal lattices. 
Note that by "chiral", we refer to a particular subclass of TRSB superconducting condensates exhibiting phase winding that breaks mirror symmetries.
Along the same lines, solutions of one-dimensional (1D) irreps that are accidentally degenerate may also form such complex superpositions. Such cases are fine tuned, but Fermi surfaces with competing nesting tendencies, for example, are prone to near-degeneracy between several symmetry-distinct superconducting instabilities and may therefore more naturally feature such accidental degenerate superpositions. Examples of these include
$d+is$ and $d+ig$ as discussed in the context of iron-based superconductors and Sr$_2$RuO$_4$ (see Fig. \ref{fig:fig0} B,C), or the so-called $s+is$ state with nontrivial phases on different bands as displayed in the phase diagram in Fig. \ref{fig:fig0} A, which also arises as a result of frustrated pairing interactions which are discussed in view of Fe-based superconductors~\cite{Stanev2010,Carlstroem2011,Maiti2013} and other multiband materials\cite{Mandal2022}. More recently, accidental degeneracies are also explored for UTe$_2$ where candidate states include complex spin-triplet superpositions of the form e.g. $A_u+iB_{3u}$, $B_{1u}+iB_{3u}$ or $B_{3u}+iB_{2u}$ now expressed directly in terms of the 1D irreps, see Fig. \ref{fig:fig0} D. All the complex superpositions mentioned above break time-reversal symmetry, as evident e.g. from the chosen directionality of the orbital motion of the Cooper pairs.

\begin{figure*}[tb]
    \centering
    \includegraphics[width=1.0\linewidth,scale=1.0]{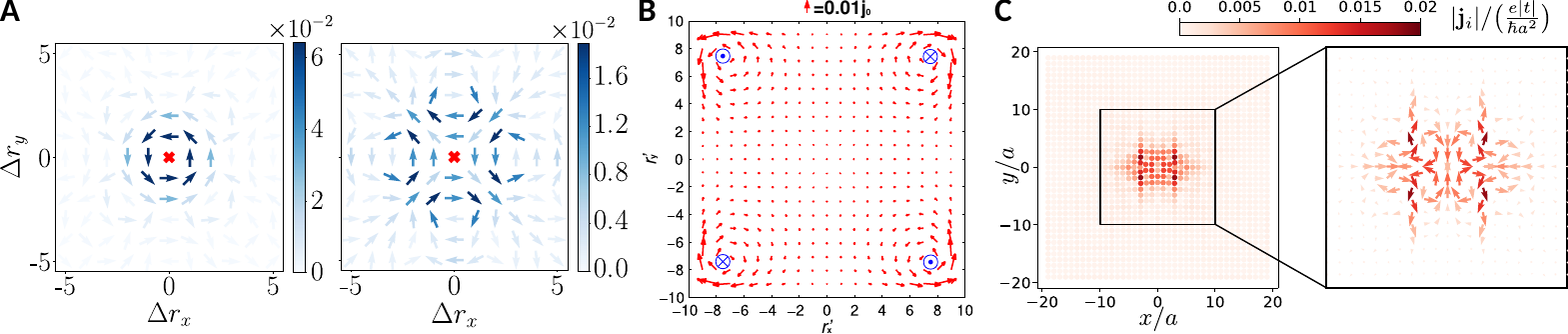}
    \caption{Examples of static supercurrent patterns induced by spatial perturbations in TRSB superconductors. {\bf A} Induced supercurrents from a single point-like nonmagnetic impurity for the cases of $d_{x^2-y^2}+id_{xy}$ (left) and $d_{x^2-y^2}+ig_{xy(x^2-y^2)}$ (right) order parameters obtained from selfconsistent studies of the extended one-band Hubbard model~\cite{Merce2022}. {\bf B} Supercurrent distribution of a square system with chiral corner loop currents in an $s+id_{x^2-y^2}$ state~\cite{Lee2009}. {\bf C} Spontaneous supercurrents induced near a dislocation modelled by five missing atoms. In this case the bulk superconductor is a pure $d_{x^2-y^2}$ state, yet the dislocation is able to locally stabilize nontrivial phase gradients with associated current patterns~\cite{Clara2023}.
    (A) Reprinted figure with permission from \cite{Merce2022}, Copyright 2022 by the American
Physical Society.
(B) Reprinted figure with permission from \cite{Lee2009}, Copyright 2009 by the American
Physical Society.
(C) Reprinted figure with permission from \cite{Clara2023}, Copyright 2023 by the American
Physical Society.}
    \label{fig:fig1}
\end{figure*}

Even though the above homogeneous superconducting condensates spontaneously break time-reversal symmetry, their unambiguous detection in actual materials is challenging experimentally since the internal magnetic fields produced in these states are suppressed due to the condensate of overlapping Cooper pairs and Meissner screening. However, defects, edges, dislocations, twin boundary interfaces, and other spatial inhomogeneities are known to produce persistent current patterns with associated magnetic fields that do not average to zero~\cite{Matsumoto1999,Stone2004,Lee2009,Garaud2014,Maiti2015,Watashige2015,Haakansson2015,Lin2016,Silaev2017,Holmvall2018,GaraudPRL,Benfenati,SongPRL2020,HolmvallBdG2020,WillaEA21,Merce2022,Clara2023}. In Fig.~\ref{fig:fig1} we show several examples of disorder- and dislocation-induced current patterns in TRSB superconducting phases. The examples include cases where the homogeneous superconducting phase breaks time-reversal symmetry (Fig.~\ref{fig:fig1} A and B), and a case (Fig.~\ref{fig:fig1} C) where TRSB is only induced locally due to the dislocation, whereas the clean homogeneous superconducting phase preserves time-reversal symmetry, see also Ref.~\cite{Stanev2014}. Naturally, a remaining important question is whether induced fields from this mechanism exhibit the required amplitudes and are voluminous enough to explain experiments of specific materials, or one needs to resort to other explanations for the TRSB signal. We return to this discussion in the final section of the paper.

\subsection{Disorder and superconductivity-induced quasi-ordered magnetism}
\label{momentmechanism}
\subsubsection{Magnetic droplets from impurities}
Nonmagnetic disorder has been known for many years to give rise to magnetic bound states and magnetic glassy behavior in correlated electron systems~\cite{Alloul2009}.  In superconductors, the opening up of a gap promotes the existence of these bound states. For fixed correlation and gap size, increasing disorder ``freezes" the spin fluctuations 
present in the system~\cite{Andersen2010}, transferring the spectral weight to low frequencies and creating static configurations of localized spins that can break time-reversal symmetry. There are several  models that exhibit these phenomena in a qualitative sense, and many somewhat mysterious aspects of the phenomenology of underdoped cuprates can be explained on this basis~\cite{Alloul2009}.  The approach has been to begin with the study of simple, localized and isolated defects in model Fermi systems with strong antiferromagnetic correlations, like Hubbard, $t-J$, and Anderson models, combined as needed with BCS pairing interactions~\cite{Tsuchiura2001,ZWang2002,Zhu2002,Chen2004,Andersen2007,Harter2007,Schmid_2010,Gastiasoro2013,Martiny2015,Martiny2019}. The magnetic response generally takes the form of a "droplet" of magnetic correlations around a defect, whose effective size corresponds to the antiferromagnetic correlation length $\xi_{\mathrm{AF}}$ of the pure host. If the correlations and/or defect potentials are sufficiently strong, this droplet is spontaneously nucleated, carries net spin $1 / 2$ and may be considered as a localized moment, but also displays a staggered magnetic behavior over $\xi_{\mathrm{AF}}$~\cite{Tsuchiura2001,ZWang2002,Zhu2002,Chen2004,Andersen2007,Harter2007}. If correlations or the impurity potential are weaker, the correlated spin response is still present, but must be induced by an applied magnetic field~\cite{Harter2007,WeiChen2009}. This is a mean-field picture of the freezing of spin fluctuations present in the host, which needs to be supplemented by a description of dynamics, a complete version of which is still lacking. In the quantum critical regime, some aspects simplify, and the basic qualitative features have been discussed in Refs.~\cite{CastroNeto1998,Millis2001}.
\begin{figure*}[tb]
    \centering
      \includegraphics[width=1.0\linewidth,scale=1.0]{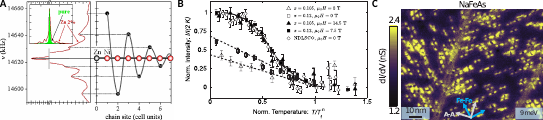}
    \caption{{\bf A} Zn impurity on 1D inorganic chain YBa$_2$NiO$_5$~\cite{Das2004}. NMR satellite lines (left) indicate different Ni nuclei at different distances from the Zn. Polarization (right) on each site oscillates in sign with distance \cite{Alloul2009}; {\bf B}
    Elastic neutron scattering data on underdoped LSCO showing the intensity at the incommensurate Bragg peaks near ($\pi,\pi$) vs. temperature $T$\cite{Chang2008}.  While a magnetic field induces SDW order, significant incommensurate magnetism is present already in zero field. This effect is enhanced with Zn substitution~\cite{Kimura2003}; {\bf C} Highly anisotropic electronic states induced by strain near a NaFeAs grain boundary, interpreted as magnetic stripes along the Fe-Fe directions (blue arrows)~\cite{Rosenthal2015_PhD}.
       (A) Reprinted figure with permission from \cite{Alloul2009}, Copyright 2009 by the American
Physical Society.
(B) Reprinted figure with permission from \cite{Chang2008}, Copyright 2008 by the American
Physical Society.
(C) Reprinted from \cite{Rosenthal2015_PhD}.}
    \label{fig:exptl_imp_states}
\end{figure*}

Away from the quantum critical point, the droplet picture has been extensively confirmed for dilute, intentionally substituted defects, primarily by NMR and $\mu$SR experiments in quasi-1D inorganic chain and ladder compounds (see Fig.~\ref{fig:exptl_imp_states} A)~\cite{Das2004}, as well as in some quasi-2D cuprate materials where impurity substitution into quite clean hosts can be adequately controlled~\cite{Alloul2009}. On the other hand, there is no reason why similar physics should not take place when defects occur naturally in the crystal or film growth process, which is characteristic of many complex oxides, particularly chemically doped ones. The question then naturally arises as to what happens when the inter-defect spacing becomes comparable to the AF correlation length $\xi_{\mathrm{AF}}$. In calculations for a 2D Hubbard model with disorder and $d$-wave pairing correlations,  the overlap
of the magnetic droplets in the correlated state leads to interference processes which tend to align the droplets in a state which displays quasi-long range AF order~\cite{Andersen2007,Andersen2010,Shender1991,Andersen2008,Christensen2011}. It seems likely that the results of many experiments which are attributed to "stripe order" in underdoped cuprates may in fact be explained by the quasi-long range order arising from interacting droplets~\cite{Schmid_2010,Schmid_2013}. The creation of 1D structures to lower the kinetic energy via 1D rivers of mobile charge while maintaining some correlation energy is clearly an attractive and simple picture. But creation of such structures also raises the energy of the state around a single impurity, which would prefer to be more globular in shape.  One can, within Hubbard-type models on square lattices, answer the question about which mechanism enables the system to lower its energy more efficiently in which situation. This clearly depends on a variety of factors, including doping, correlations, and disorder. By performing simulations, one can - without prejudicing the system - determine which part of the generalized phase diagram corresponds to 1D stripe-like structures, and which to other ground states~\cite{Kivelson2003,Vojta2009,Schmid_2010}. 

Experimentally, manifestations of this physics have been observed in a wide variety of contexts.  In Fig.~\ref{fig:exptl_imp_states} A, we show the well-established oscillations in the Y NMR line in a 1D inorganic chain due to the modulation of the magnetic response locally by a Zn impurity. 
Figure~\ref{fig:exptl_imp_states} B shows data from an elastic neutron study~\cite{Chang2008} (see also \cite{Lake2002,Khaykovich2002,Romer2013}) with frozen magnetic order in a disordered cuprate in zero field, enhanced further by application of a field. The order is apparently quasi-long-ranged since it appears in the neutron scattering intensity near wave vector $(\pi,\pi)$, yet it occurs in a part of the cuprate phase diagram where true long-range order has disappeared.  Finally, in Fig.~\ref{fig:exptl_imp_states} C, STM reveals in an Fe-based superconductor a spatially modulated density of states pattern due to strain near a twin boundary with a wave vector close the $(\pi,0)$ magnetic ordering vector, indicating local impurity-induced magnetism. 

\subsubsection{Quasi-long range magnetic order from disorder}

\begin{figure*}[tb]
    \centering
                \includegraphics[width=1.0\linewidth,scale=1.0]{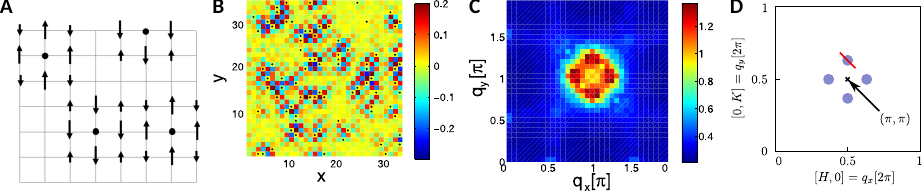}
    \caption{{\bf A} Schematic of magnetic droplets created around individual nonmagnetic impurities in a correlated system, which interact via an effective collective exchange, which may be understood by the need to maintain Néel coherence among the different droplets. {\bf B}, typical magnetization map $M(\bf{r})$ of a disordered superconductor is shown for $U=3.2t$~\cite{Andersen2007}. {\bf C} the Fourier transform of $M(\bf{r})$ yields the structure factor $S(\mathbf{q})$.  {\bf D} Sketch of the expected intensities (light blue dots) away from $(\pi,\pi)$ (black cross) as measured in neutron scattering experiments by taking data along line cuts (red line)~\cite{Lake2002}.
           (A, B, C) Reprinted figure with permission from \cite{Andersen2007}, Copyright 2007 by the American Physical Society.}
    \label{fig:order_by_disorder}
\end{figure*}

Quasi {long-range antiferromagnetic order} has been reported in many cuprate samples whose $T_{\mathrm{c}}$ or nominal doping puts them  outside the long-range AF phase, particularly  in the superconducting state at low
temperatures, e.g. in
LSCO~\cite{Lake2002,Kimura2003,Watanabe2002,Chang2008,Khaykovich2002,Khaykovich2005,Niedermayer1998,Panagopoulos2002}.
Characteristic dynamical features of spin glasses have occasionally been observed, but even if these are not always found, the term ``spin glass"  has become commonplace in cuprates to describe the existence of a phase of short-range frozen magnetism. The size of the spin-glass phase varies significantly with material~\cite{Panagopoulos2002};  whereas in LSCO it extends in zero field out to 15\% doping, it is confined to very low doping in the much cleaner YBCO system~\cite{Miller2006,Haug2009}, and BSCCO is in-between~\cite{Panagopoulos2002}. Moreover, disorder due to Zn substitution on the Cu sites enhances the size of the spin-glass phase~\cite{Kimura2003}.  
These empirical facts suggest strongly the picture of disorder-induced short-range magnetism described above.   Quasi-long range  order
occurs when these droplets begin to overlap~\cite{Andersen2007}, which can occur without frustration on a square lattice, as shown in Fig.~\ref{fig:order_by_disorder} A. This lack of frustration has important implications not only for the quasi-long range ($\pi,\pi$) magnetic order, but may also carry a macroscopic net magnetic moment.  Since each droplet of {\it staggered} order is dominated by the NN spins, it carries a net spin 1/2~\cite{ZWang2002}, and the droplets overlap coherently.  While the net moments associated with each droplet do not align ferromagnetically, a fluctuating magnetic state with both short range and longer range (inter-impurity distance) is created (Fig. \ref{fig:order_by_disorder} B). The final magnetization profile appears as a clear peak in the structure factor, as seen in Fig.~\ref{fig:order_by_disorder} C, compared to the Lake {\it et al.} experiment on underdoped LSCO where intensity at the incommensurate positions is found~\cite{Lake2002}, as shown schematically in Fig. \ref{fig:order_by_disorder} D. It is this disorder-induced random magnetic landscape that is relevant to our current study as a possible mechanism of TRSB in superconductors.
  \begin{figure*}[tb]
    \centering
        \includegraphics[width=\linewidth,scale=1.0]{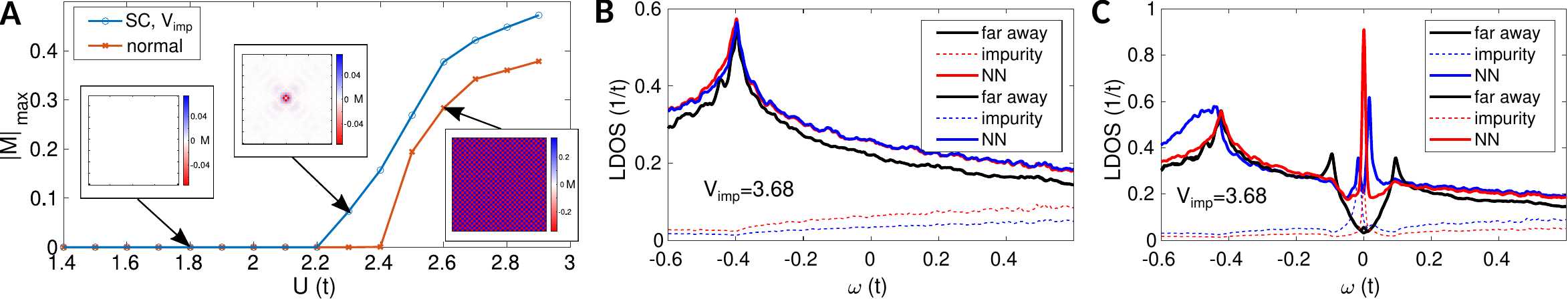}
    \caption{{\bf A} Magnetic phase diagram for a superconductor including a nonmagnetic impurity and a normal correlated metal. {\bf B} Density of states of a normal  metal including a nonmagnetic impurity with correlations (blue curves, $U=2.3t$) and without correlations (red curves, $U=0$). {\bf C} Density of states of a $d$-wave superconductor including a nonmagnetic impurity with and without correlations.}
    \label{fig:bound_st}
\end{figure*}

\begin{figure*}[tb]
    \centering
         \includegraphics[width=0.9\linewidth,scale=1.0]{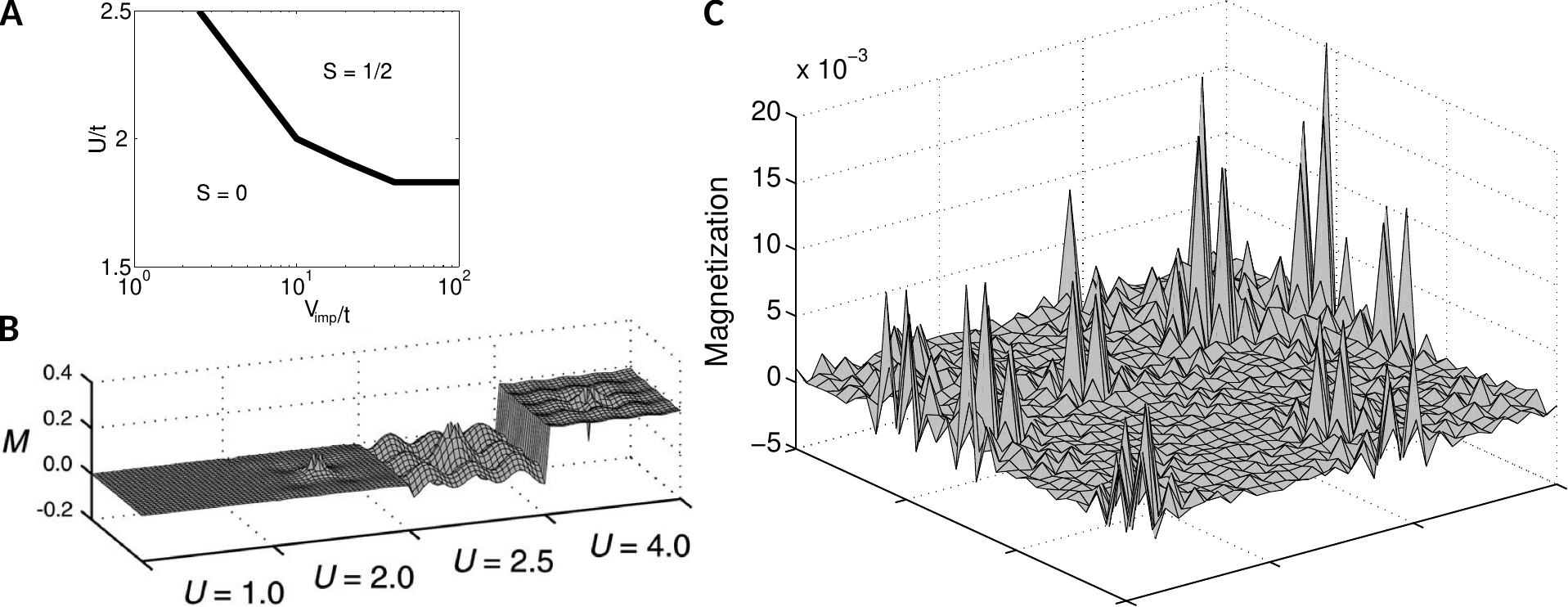}
    \caption{{\bf A} phase diagram for single impurity in $d$-wave superconductors.  $U$ is the Hubbard interaction, treated in Hartree-Fock approximation, and $V_{\rm imp}$ is the impurity potential. {\bf B}: staggered magnetization near a strong impurity for increasing correlation strength $U$. {\bf C}: magnetization landscape induced by a weak magnetic field in the presence of many impurities without spontaneous spin freezing. From Ref. \cite{Harter2007}.
    (A,B,C) Reprinted figure with permission from \cite{Harter2007}, Copyright 2007 by the American Physical Society.}
    \label{fig:Harter_1imp}
\end{figure*}

Thus far we have said little about why such a TRSB signal might arise only below $T_{\mathrm{c}}$, as observed in many unconventional superconductors.  Here the proposal is that the gapping of the low-energy density of states enhances the probability of creating an impurity bound state carrying the spin-1/2 moment. Figure~\ref{fig:bound_st} A shows the selfconsistently obtained magnetization in the presence of a nonmagnetic point-like impurity at $T=0$ versus Hubbard $U$. As seen, before entering the bulk SDW phase, a regime exists where the superconductor prefers a local impurity-induced magnetic droplet as discussed above. What is the origin of this local magnetic structure? In Fig.~\ref{fig:bound_st} B, we show the local density of states of the tight-binding model with band parameters suitable for a cuprate superconductor.  Addition of a strong impurity does not produce a bound state resonance due to the fact that the impurity state is embedded in the continuum.  Addition of correlations does not change this qualitative picture.  On the other hand, when a superconducting gap is created, as in the $d$-wave case shown in Fig.~\ref{fig:bound_st} C, the expected in-gap resonance is clearly visible (compare, e.g. with Zn impurity in BSCCO~\cite{Pan2000}).  In the presence of correlations, the bound state splits to lower the overall energy including magnetism, creating in the process the localized spin-1/2 state with maximal polarization typically on nearest neighbor sites.
Creating such a state depends both on the impurity potential $V_{\rm imp}$ and on the strength of the correlations $U$, as shown in the ``phase diagram" of Fig.~\ref{fig:Harter_1imp}A.  The explicit evolution of the 1-impurity local magnetic state for increasing correlation strength is shown in Fig.~\ref{fig:Harter_1imp} B: it is stabilized in a range of $U$ just below the critical $U$ corresponding to creation of true long-range order in the Hartree-Fock approximation.  Even if the magnetic bound states are not formed spontaneously, i.e. one is in the $S=0$ part of the phase diagram shown in \ref{fig:Harter_1imp} A, it is important to recall that the system is still highly correlated as indicated by the response of many impurities for small but nonzero $U$ to a weak magnetic field, as seen from Fig.~\ref{fig:Harter_1imp} C.  

Similar phenomena have been observed in many other correlated superconductors.  Most recently, magnetic clustering was reported by $\mu$SR in samples of UTe$_2$ grown by chemical vapor transport with $T_{\mathrm{c}}$'s of order 1.6K~\cite{Sonier2023A}, whereas no magnetic signal was reported in cleaner molten-salt flux grown crystals with $T_{\mathrm{c}}$'s of order 2.1K~\cite{Azari2023}.

To summarize this section, simulations of a $d$-wave superconductor in the presence of local Coulomb interactions and
nonmagnetic disorder~\cite{Andersen2007,Andersen2008,Schmid_2010,Schmid_2013} are compatible with the experiments on relatively disordered cuprates like LSCO and
BSCCO where a spin-glass phase is observed at low doping in the SC
state~\cite{Panagopoulos2002}. These studies explain as well why such a  phase is {\it not} seen in the cleanest YBCO samples, why in LSCO and BSCCO the spin-glass phase is more robust at
underdoping, and why it is enhanced by addition of further disorder.
The order-from-disorder many-impurity magnetic state is favored to appear only below $T_{\mathrm{c}}$ when the gap opens. Similar phenomena have been observed in other correlated superconductors~\cite{Inosov2013,GastiasoroPRL2014,Urbano2007,Martiny2015}.

\subsubsection{TRSB from disorder-induced loop currents}

\begin{figure*}[tb]
    \centering
        \includegraphics[width=\linewidth,scale=1.0]{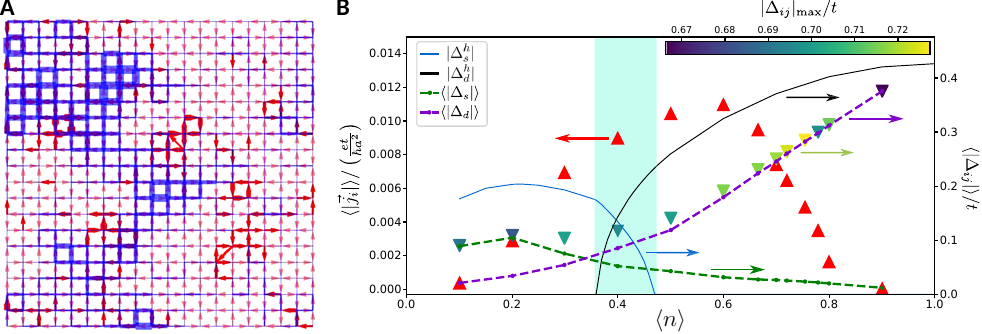}
    \caption{{\bf A}  Spontaneous current pattern in a $d$-wave superconductor with large disorder \cite{Li2021}. {\bf B} Zero temperature phase diagram indicating average current (red triangles, left scale) vs. doping.  Also indicated are  $s$- and $d$-wave order parameter amplitudes, respectively (solid lines). The shaded region marks the range of doping where a mixed state $s+id$ is stable in the homogeneous case. Dashed lines show the average $d$-wave (purple) and $s$-wave (green) order parameter components (see main text) with random distributions of impurities \cite{Clara2022}.
    (A) Reproduced from \cite{Li2021}. CC BY 4.0.
    (B) Reprinted figure with permission from \cite{Clara2022}, Copyright 2007 by the American Physical Society.}
    \label{fig:currents}
\end{figure*}

Recently, a Bogoliubov-de Gennes  study of a disordered $d$-wave superconductor without magnetic correlations~\cite{Li2021} reported local TRSB current loops for sufficiently high disorder levels, see Fig.~\ref{fig:currents} A. Since a pure $d$-wave superconducting state does not break time-reversal symmetry, and parameters are chosen to place the pure system far from TRSB mixed-symmetry states ($d+i s$), it is surprising that such persistent currents can arise purely from nonmagnetic disorder. 
In addition, the possible emergence of an effective granular $d$-wave description was proposed by Li {\it et al.}~\cite{Li2021}.  In this picture, cuprates consist of intrinsically inhomogeneous $d$-wave regions of roughly constant phase, weakly coupled by Josephson tunneling. For example, such behavior has been claimed experimentally from STM, magnetization and transport measurements on certain overdoped LSCO samples~\cite{Tranquada2022}. Such a proposal, if correct, would have profound implications for the disappearance of superconductivity on the overdoped side of the superconducting dome, since phase fluctuations, rather than pair interaction weakening (as due, for example, to the weakening of spin fluctuations and disorder~\cite{Maier2020}) would determine $T_{\mathrm{c}}$.

However,  detailed theoretical investigations of the disorder-induced orbital currents in the same model framework were performed in Ref.~\cite{Clara2022},  showing
that the occurrence of such currents can be traced to local extended $s$-wave pairing out of phase with $d$-wave order, despite the fact that the effect occurs far from the homogeneous $s+id$ phase (see Fig. \ref{fig:currents} B).  The system does not appear to spontaneously phase separate into granular regions.  Instead, the energetics leading to regions of {\it local} $s \pm i d$ order are driven by inhomogeneous density modulations. 

\subsection{Enhanced RKKY exchange coupling and slowing-down of magnetic fluctuations by superconductivity}
\label{couplingmechanism}

While the previous sections addressed the case where superconductivity is a necessary ingredient for orbital currents or moment formation, here we focus on different scenarios where disorder and superconductivity interact to either slow down magnetic fluctuations or enhance the mutual coupling between magnetic regions/moments existing already in the normal state. This section serves to suggest alternative mechanisms for observing TRSB below $T_{\mathrm{c}}$. We focus on the physical picture and outline several options whereas a quantitative study is left for future research. We consider the following possibilities:
\begin{itemize}
    \item (A) Time-reversal symmetry is preserved in the normal state, but disorder pins or slows down fluctuations in the superconducting state to either break time-reversal symmetry or bring the system closer to local magnetism.
    \item (B) The normal state contains significant amounts of magnetic moments, but time-reversal symmetry remains preserved in the normal state. The magnetic moments get resonantly coupled in the superconducting state thereby breaking time-reversal symmetry or bringing the system closer to static magnetism.
    \item (C) Time-reversal symmetry is broken already in the normal state, yet too weak or dilute to be detected by standard experimental probes. Superconductivity enhances the magnetic volume fraction and/or increases the coupling between magnetic regions.
\end{itemize}

The first {\bf scenario (A)} was largely discussed in Sec.~\ref{momentmechanism}, there exemplified through models relevant for cuprates, i.e. $d_{x^2-y^2}$-wave superconductivity in the presence of nonmagnetic impurities and electronic correlations. This mechanism is likely much more broadly applicable since most unconventional superconductors allow for in-gap bound states from nonmagnetic disorder~\cite{BalatskyRMP,Gastiasoro2013,Holbaek2023}. This causes enhanced local spin susceptibilities and associated tendency to slow down magnetic fluctuations locally or even pin magnetic moments~\cite{Alloul2009}. This mechanism may well also be at play near e.g. sample edges or various forms of dislocations~\cite{WillaEA21,Clara2023,Pal2021,Pal2023}. A related property may happen near sample edges when zero-energy Andreev bound states generated by sign-changes of the gap function split by spontaneous generation of edge supercurrents~\cite{Haakansson2015,Holmvall2018,HolmvallBdG2020}.

Regarding {\bf scenario (B)}, consider the case of a normal metal with a concentration of magnetic moments. Due to quantum fluctuations, time-reversal symmetry is preserved in the normal state. We assume additionally a low Kondo temperature, $T_{\mathrm{K}}<T_{\mathrm{c}}$, i.e. unscreened spins. We return briefly to the opposite case with $T_{\mathrm{K}}>T_{\mathrm{c}}$ further below. The magnetic moments couple via the indirect Ruderman-Kittel-Kasuya-Yosida (RKKY) interaction, i.e. via a partial spin polarization of the conduction electrons, causing the well-known oscillatory RKKY exchange interaction~\cite{RudermanKittel,Kasuya,Yosida}. The detailed behavior of the RKKY coupling in the superconducting phase depends on dimensionality, the normal state bandstructure, and the superconducting pairing symmetry, and is generally a complex matter that requires thorough investigation~\cite{abrikosov1988fundamentals,Aristov1997,Galitski2002,Schecter2016}. Typically, however, for spin-singlet superconductivity the RKKY coupling is weakened by the singlet formation of the conduction electrons. More specifically, for the standard RKKY interaction $s$-wave superconductivity contributes a weak antiferromagnetic correction when the inter-moment distance $R$ is smaller than the coherence length $\xi$. For $R>\xi$, the RKKY coupling becomes purely antiferromagnetic from the anomalous (superconducting) part of the spin susceptibility, but is exponentially suppressed in $R$~\cite{abrikosov1988fundamentals}. These are consequences of the typical suppressed spin susceptibility in the superconducting phase. It is important to emphasize that these properties apply to the perturbative RKKY expression calculated in the Born approximation, only in the case of $s$-wave spin-singlet superconductivity.

It is known, however, that another contribution to the exchange coupling arises from the Yu-Shiba-Rusinov (YSR) bound states produced by the magnetic impurities~\cite{Yu,Shiba,Rusinov,BalatskyRMP,Yao2014}. The overlap of the YSR bound state wavefunctions leads to a resonant-like  contribution to the energy with a singular dependence of the exchange coupling on the YSR bound state energy~\cite{Yao2014}. Thus, in some superconductors where the average distance between the magnetic moments is lower than $\xi$, it is conceivable (though somewhat fine-tuned due to the requirement of very low-lying YSR states~\cite{Yao2014,Hoffman2015}) that the RKKY coupling gets enhanced in the superconducting phase, thereby boosting the tendency for magnetism and the detection of a TRSB signal only below $T_{\mathrm{c}}$. In fact, this mechanism may also be at play for the mechanism discussed in Sec.~\ref{momentmechanism} where magnetic moments are generated only by electronic interactions and superconducting bound states from nonmagnetic impurity potentials. To the best of our knowledge, the resonant-like YSR energy contribution has not been thoroughly investigated for other superconducting pairing symmetries, but presumably nodal gaps or triplet order can drive even stronger, directional-dependent couplings~\cite{Aristov1997}.  Quantum corrections to the classical YSR picture can also in principle lead to TRSB enhancement in the superconducting state, if in the case $T_{\mathrm{K}} \gtrsim T_{\mathrm{c}}$, spin-singlet superconducting gap formation unscreens
the moments. In that case, if the moments couple and generate a weak magnetic state, it may be picked up by e.g. $\mu$SR similarly to the scenario discussed in Sec.~\ref{momentmechanism}. 

Finally we discuss {\bf scenario (C)} where a distribution of disordered static magnetic moments are already present in the normal state.
Thus in this case, strictly speaking, TRSB is not restricted to the superconducting phase below $T_{\mathrm{c}}$.
We do not discuss in detail the origin of such moments, but they could simply arise from weakly coupled magnetic impurities, or small spin-polarized regions that have crossed into a nearby magnetic phase due to chemical disorder. It might also be due to the mechanism discussed in Sec.~\ref{momentmechanism} if the "normal state" exhibits pseudo-gap behavior. We assume that the magnetism in the normal state is too weak to be detected by standard probes, for example due to a tiny volume fraction and/or short-range nature. A TRSB signal may be detected, however, in cases where the volume fraction grows and/or the moments couple more strongly, and the magnetic quasi-order becomes longer ranged as a result upon entering the superconducting phase~\cite{Fittipaldi2021,Sonier2023A}. The latter may arise from the mechanism discussed under (B). It might also take place via new "stepping stone" magnetic islands produced by nonmagnetic disorder in the superconducting state via the mechanism discussed in Sec.~\ref{momentmechanism}. An enhanced volume fraction can take place in cases where another competing order is in the game. If local disorder-induced magnetism is limited by orbital order or a CDW phase, for example, then superconductivity may suppress these competing orders and thereby alleviate the magnetism. Clearly, these are merely possibilities for observing TRSB below $T_{\mathrm{c}}$, and need further detailed investigations for specific materials.

\section{Experimental consequences}

In this section we discuss the main current experimental techniques to directly probe TRSB in superconductors, and mention subtleties in the interpretation of experimental results. Many unconventional superconductors are investigated using the muon spin relaxation ($\mu$SR) technique where a beam of spin-polarized muons with given energy is directed to the sample~\cite{blundell2022muon} . Muons are stopped in the material, implanted at preferred positions in the crystal structure and can be used to detect the magnetic field locally at their positions. Within the lifetime of the muons, their spins precess in the local field. Once the muon decays into a positron and two neutrinos, the direction of the magnetic moment fixes the probability distribution of emitted positrons. The counting rates of the latter in the forward and backward directions of the (original) polarization of the muon beam yield the asymmetry $A(t)$, which is recorded as a function of time.

The Fourier transform of $A(t)$ for measurements on superconductors in a finite field (in the vortex state) is directly related to the magnetic field distribution of the sample and can be used to deduce the temperature dependence of the penetration depth~\cite{blundell2022muon,Sonier2000,Khasanov2015}. In a zero field experiment, the information about the field distribution is also contained in the asymmetry function, but typically full oscillations are not observed in superconductors because of the smallness of the internal field. Still, an increased relaxation as a function of temperature is taken as signature of the occurrence of an additional internal field that collectively depolarizes the implanted muons.  Average fields that are static on the time scale of the muon decay can then be detected down to the scale of $\mu$T~\cite{blundell2022muon}.
From the shape of the asymmetry $A(t)$ at low temperature (when there is evidence for TRSB), it is possible to estimate the average field by the reduction at largest measuring times. A quick initial reduction and no further deviation as a function of time are indications of few muons getting depolarized in a large field. By contrast, an initial weak difference of the depolarization rates that behaves quadratically in time indicates an almost uniform field distribution. Fitting to such behavior can give bounds on the fraction of the sample exhibiting TRSB. Such bounds may be further benchmarked against scanning SQUID microscopy measurement~\cite{HicksEA10,Curran2023,Mueller2023}. Depending on the initial energy of the muons, implantation happens on length scales of 10\,nm to 100\,nm from the surface, such that the technique can be considered as bulk-like. By varying the initial energy of the muons, the stopping profiles allow deductions of the TRSB as function of distance from the surface~\cite{Fittipaldi2021}. The $\mu$SR technique has been used to claim TRSB in a number of superconductors; for a list see Ref.~\cite{Ghosh2EA20}. 

Knowledge from {\it ab initio} methods about the preferred position within the elementary cell where muons are implanted~\cite{Huddart2021,Blundell2023} can facilitate the interpretation of the results, i.e. help decipher whether the positive charge of the muons might cause a perturbation large enough to self-generate a TRSB signal. In addition, the position of the stopping site and the associated form factor can be important for a quantitative description of the relaxation, e.g. in cases where the muon locations may feature enhanced or reduced field effects from purely geometric reasons.  The typical magnitude of the internal field at the muon site for a generic interstitial position in the unit cell may be  estimated;  in the vicinity of local spin-1/2 droplets (Sec.~2.2) or impurity-induced localized orbital currents are of order 1~G ~\cite{Ghosh2EA20,Miyake2017},  which is of the same order as fields reported in experiments.

The second main probe of TRSB is the polar Kerr effect, where the polarization plane of light incident on a sample is rotated upon reflection. The classic example is reflection from a ferromagnetic surface, but in recent years the technique has been employed to detect TRSB in unconventional superconductors~\cite{Wysokinski2019,Kallin16,SCHEMM201713,HayesEA21,Ajeesh2023,XiaEA06}. The angle $\theta_{\mathrm{K}}$, which is related to the Hall conductivity, can be expressed as
$$
\theta_{\mathrm{K}}(\omega)=\frac{4 \pi}{\omega} \operatorname{Im}\left(\frac{\sigma_{\mathrm{H}}(\omega)}{n\left(n^2-1\right)}\right),
$$
where $n$ is the complex, frequency-dependent index of refraction, and $\sigma_{\mathrm{H}}(\omega)$ is the Hall conductivity.
For the experimental identification of a polar Kerr effect arising from the superconducting state, it needs to be checked that the rotation onsets at $T_{\mathrm{c}}$, the sign and magnitude of the signals vary after each  cool-down in the absence of an external magnetic field, a small training field saturates the rotation angle (but with sign given by the direction of the field), and the temperature dependence of the Kerr angle follows mean-field-like order parameter behavior~\cite{SCHEMM201713}.

From a theoretical point of view, a finite Kerr angle is directly related to a nonzero Hall conductivity, which requires time-reversal symmetry breaking and breaking of mirror symmetries about all planes parallel to the incident wave vector~\cite{Cho2016}.
Therefore, the order parameter has to be chiral and not simply break TRSB.  For example, a pure $s+id$ superconductor is nonchiral and will not generate an intrinsic Kerr signal.  By contrast, a one-band $p+ip$ superconductor will produce a weak intrinsic Kerr effect with  $\theta_{\mathrm{K}}$  arising only from the collective modes of the order parameter~\cite{YipSauls1992A}. This effect is many orders of magnitude smaller than typical measured rotation angles $\theta_{\mathrm{K}}$ of order tens of nanoradians.   The intrinsic effect in a chiral {\it multiband} superconductor, arising from contributions to $\sigma_{\mathrm{H}}$ from interband transitions, is larger at optical frequencies (but still difficult to definitively quantify)~\cite{Taylor2012}.  On the other hand, if the state is nonchiral, there will be no intrinsic Kerr signal at all.

It is important to note that chirality can occur for complex linear combinations of 1D irreps of the point group, and not necessarily only those states corresponding to 2D irreps that have been traditionally discussed, like $p_x+ip_y$ in tetragonal and $d_{xy}+id_{x^2-y^2}$ in hexagonal symmetries.  For example, in tetragonal symmetry, $d_{x^2-y^2}$ and $d_{xy}$ are distinct 1D irreps, but may occur in an accidental linear combination $a\cdot d_{xy}+i b\cdot d_{x^2-y^2}$.  Such states should also give rise to an intrinsic polar Kerr effect, but nonchiral TRSB combinations like $s+id$ will not.

One of the problems with the interpretation of Kerr measurements is that the size of an intrinsic signal is very hard to estimate, and the size of an extrinsic one may in principle be in fact much larger, but depend on the type of disorder and the nature of the ground state. For example, 
the division between chiral and non-chiral states with regard to their polar Kerr signals is muddied somewhat by disorder. In $p+ip$ states, an extrinsic  Kerr rotation angle orders of magnitude larger than the intrinsic effect may  result from the skew-scattering of chiral Cooper pairs from impurities~\cite{Goryo2008,Lutchyn2009}, but for $d+id$ and higher order angular momentum states, the contribution is of higher order in $(nV_{\rm imp}^2)$ where $V_{\rm imp}$ is the impurity potential of point-like impurities with density $n$~\cite{Kallin16,Goryo2008}.  Estimates in both cases are model dependent and require disorder averaging.   Impurities in nonchiral TRSB states will also give rise to small signals from induced currents, but after averaging such signals should be substantially smaller, while for chiral $d+id$ states the signal is small but remains finite upon impurity averaging~\cite{Liu2023}.  A similar contribution should therefore be provided by local spatially fluctuating magnetic moments.  Whether or not a Kerr signal is observed, and its size, depends on how much averaging of the fluctuations takes place over the laser spot.  In both these cases, some kind of ability to ''train`` the state with an external magnetic field should be observed~\cite{SCHEMM201713}.  Finally, we mention some recent calculations showing that local currents can be created near dislocations, which induce $d+ig$ or $s+id$  states~\cite{WillaEA21,Clara2023}. The Kerr effect arising from these local states has not been properly explored.

\section{Summary and Outlook}

In the above we focused on different theoretical mechanisms where magnetism in various guises appear only below the superconducting critical transition temperature $T_{\mathrm{c}}$ and is generated by disorder or other spatial inhomogeneities. This included discussions of localized defect-induced orbital supercurrents, magnetic moment generation from superconductivity and electronic correlations, and the possibility of enhanced RKKY exchange coupling by superconductivity.

Naturally an important question relates to the relevance of these mechanisms to the observation of TRSB below $T_{\mathrm{c}}$ in currently discussed unconventional superconducting materials. In the above, we provided a detailed discussion of the cuprates and the possibility of superconductivity-enhanced spin-polarization. Two other materials of significant current interest are UTe$_2$ and Sr$_2$RuO$_4$. For these compounds, two-component condensates are widely discussed, but inconsistent with specific heat measurements, also under uniaxial strain featuring no signs of double transitions~\cite{SteppkeEA17,LiEA19,RosaEA22,GirodEA22}. This points to single-component superconductivity which has received recent support from a number of other probes as well~\cite{GrgurEA23,Ajeesh2023,Azari2023,Theuss2023}, but leaves open the question of what causes a TRSB signal in some measurements. We have surveyed several possible answers to this question from the perspective of induced fields within the superconducting state from spatial inhomogeneity.

Resolving the detailed nature of TRSB in  superconductors is important since the resolution will have profound implications for our general understanding of unconventional superconductivity. Therefore, we hope that future collective research efforts will continue to be devoted to this task. This includes both improved experiments and better theoretical descriptions, the latter including material-specific studies of the interplay between unconventional pairing states and realistic models for the relevant sources of spatial inhomogeneity.
$\mu$SR and Kerr experiments on systematically disordered samples,  e.g. with strong chemical impurities or  Frenkel defects created  by electron irradiation, can help to distinguish the extrinsic TRSB effects described here from intrinsic effects, and from each other.

\section*{Conflict of Interest Statement}

The authors declare that the research was conducted in the absence of any commercial or financial relationships that could be construed as a potential conflict of interest.

\section*{Author Contributions}

All authors contributed to writing the manuscript.

\section*{Funding}
 A.K. acknowledges support by the Danish National Committee for Research Infrastructure (NUFI) through the ESS-Lighthouse Q-MAT. P.J.H. acknowledges support from NSF-DMR-2231821.

\section*{Acknowledgments}
We acknowledge useful discussions with C. N. Brei\o, M. H. Christensen, P. Hedeg\aa{}rd, C. Hicks, M. Roig, H. R\o{}ising, A. T. R\o{}mer, and J. Schmalian. We thank M. Pal for assistance with calculations for Fig. 5.

\section*{Data Availability Statement}
The raw data supporting the conclusion of this article will be
made available by the authors, without undue reservation.
% Please see the availability of data guidelines for more information, at https://www.frontiersin.org/about/author-guidelines#AvailabilityofData

%\bibliographystyle{frontiersinSCNS_ENG_HUMS} % for Science, Engineering and Humanities and Social Sciences articles, for Humanities and Social Sciences articles please include page numbers in the in-text citations
\bibliographystyle{frontiersinHLTH&FPHY} % for Health, Physics and Mathematics articles
\bibliography{references}

\end{document}